\documentclass{llncs}
\usepackage{mathptmx}       % selects Times Roman as basic font
\usepackage{helvet}         % selects Helvetica as sans-serif font
\usepackage{courier}        % selects Courier as typewriter font
\usepackage{type1cm}        % activate if the above 3 fonts are not available on your system
\usepackage{makeidx}         % allows index generation
\usepackage{graphicx}        % standard LaTeX graphics tool when including figure files
\usepackage{multicol}        % used for the two-column index
\usepackage[bottom]{footmisc}% places footnotes at page bottom
\usepackage{subfigure}
\usepackage{amsfonts}
\usepackage[cmex10]{amsmath}

\usepackage{tabularx}
\usepackage{comment}

\usepackage[utf8]{inputenc}
\usepackage{graphicx}
\usepackage{multirow}
\usepackage{hyperref}
\hypersetup{
    colorlinks=true,
    linkcolor = {red},
    citecolor = {cyan},
    filecolor = {cyan},
    menucolor = {magenta},
    runcolor = {cyan},
    urlcolor = {blue},
}
\usepackage{amssymb}% http://ctan.org/pkg/amssymb
\usepackage{pifont}
\usepackage{subfigure}
\graphicspath{{Images/}}
\usepackage{enumitem}
\usepackage[table,xcdraw]{xcolor}

\usepackage{pifont}
\usepackage{amsmath}
\usepackage{amsfonts}
\usepackage{lmodern}

\begin{document}

\title{Network Classification and Categorization}

\titlerunning{Network Classification}  
\author{James P. Canning\inst{2}%\inst{1} 
\and Emma E. Ingram\inst{3}%\inst{1}
\and Sammantha Nowak-Wolff\inst{1} \and 
\\
Adriana M. Ortiz\inst{4}%\inst{1} 
\and Nesreen K. Ahmed\inst{5} \and 
Ryan A. Rossi\inst{6} 
\and 
\\
Karl R. B. Schmitt\inst{1} \and Sucheta Soundarajan\inst{7}
}
\authorrunning{Canning et al.} 
\tocauthor{James P. Canning, Emma E. Ingram, Sammantha Nowak-Wolff, Adriana M. Ortiz, Nesreen K. Ahmed, Ryan Rossi, Karl R. B. Schmitt, Sucheta Soundarajan}

\institute{Valparaiso University, Valparaiso IN 46383, USA,\\
    Corresponding Author: \email{karl.schmitt@valpo.edu} 
    \and
    SUNY Geneseo, Geneseo NY 14454, USA,
    \and
    University of Alabama, Tuscaloosa AL 35487, USA, \\
    \and
    University of Puerto Rico, Rio Piedras, San Juan PR 00936, \\
    \and
    Intel Labs, 3065 Bowers Ave, Santa Clara, CA USA \\
    \and 
    Xerox PARC, 3333 Coyote Hill Rd, Palo Alto, CA USA \\
    \and
    Syracuse University, 223 Link Hall, Syracuse, NY USA \\
}

\date{}

\maketitle

\vspace{-0.15in}
\section{Introduction}\label{sec:intro}
\vspace{-0.11in}
Networks are often categorized according to the underlying phenomena that they represent, such as re-tweets, protein interactions, or web page links. It is generally believed that networks from different categories have inherently unique network characteristics. 
In this work, we find strong evidence supporting this hypothesis by learning a classification model $f \;:\;\mathrm{\bf x}\, \rightarrow\, y\,$ that is able to \emph{accurately} predict (with 94.2\% accuracy) the category of a new arbitrary unknown network $G^{\prime}$ described only by a ${ D}$-dimensional feature vector ${\bf \rm x}^{\prime}$ where $y \in \{1,2,\ldots,{K}\}$ is the class label (category).
The classifier $f$ is learned using over N=$500$ networks from K=$8$ categories (See Figure~\ref{fig:classification-results}) which are characterized using only D=$15$ simple structural features (Table~\ref{tab:features}).
As an aside, Graphlet features~\cite{nesreen_graphlet} and other more discriminative features can be used to further improve the accuracy.  
    
To the best of our knowledge, this work is the first large-scale study that tests whether network categories are distinguishable from one another (using both categories of real-world networks and synthetic graphs).
Previous research has focused on either (i) classification of synthetic graphs or (ii) graphs within a particular category such as molecular graphs. Other examples include distinguishing between brain or breast cancer cells~\cite{li} or distinguishing between different social structures~\cite{ugander2013subgraph}.

A classification accuracy of 94.2\% was achieved using a random forest classifier with both real and synthetic networks.
These results indicate that while some of the categories researchers use to label their graphs are indeed distinct, others, from a feature standpoint, are largely indistinguishable from one another.
Moreover, from a feature standpoint, synthetic graphs are trivial to classify as they are structurally distinct from all other graphs. Additionally, the classifiers also highlighted networks that are outliers within their own categories, suggesting new potential directions for understanding those networks.

\vspace{-0.1in}   
\section{Data}
\vspace{-0.1in}
Data was originally pulled from the Network Repository~\cite{nr} for all non-synthetic graphs. This included 1241 graphs with 15 network features. The features in the data are listed in Table \ref{tab:features}. Of the 20 network categories included, three were from computational and algorithmic challenges (DIMACS, DIMACS10 and BHOSLIB) and two recorded graphs over time (Temporal Reachability, Dynamic Networks). As all five of these categories are fundamentally different from static one-time recorded networks from a discipline or field they were discarded as outside the problem scope. Within the 15 remaining categories, 9 categories had less than 20 instances and thus were also excluded as having insufficient data for training.
Finally, Cheminformatics had significantly more instances than all other categories and therefore was downsampled to 119 networks which is comparable to the 2$^{nd}$ largest category. We also generated 125 graphs: 50 using the Barabasi-Albert (BA) model and 75 using the Erd\H{o}s-R\'enyi (ER) model. The final classification data set has 529 graphs from 8 categories.
\vspace{-0.2in}
\begin{table}[h]
\begin{tabularx}{1.0 \linewidth}
{XXlX}
Number of Nodes\hfill & \quad Avg. Degree    \hfill & Avg. Clustering Coefficient  \hfill  & \quad Assortativity    \\
Number of Edges & \quad Min. Degree    & Fraction of Closed Triangles   & \quad Total Triangles  \\
Maximum K-core  & \quad Max. Degree    & Max. Clique (lower bound)   & \quad Avg. Triangles   \\
Chromatic Num. & \quad Density  & Maximum Triangles
\end{tabularx}
\vspace{1mm}
\caption{Features calculated by the Network Repository}
\label{tab:features}
\end{table}

\normalsize
\vspace{-0.5in}

\vspace{-0.1in}
\section{Results}
\begin{figure}[hb!]
    \vspace{-0.1in}
    \centering
    \includegraphics[width = 0.65\textwidth]{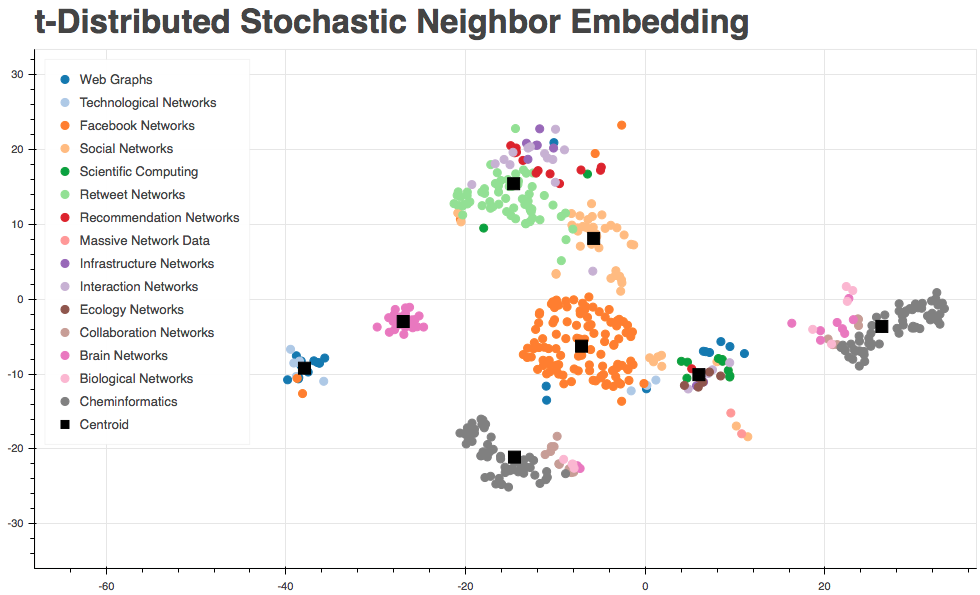}
    \caption{t-SNE Clustering. Black Squares indicate centroids from K-means}
    \label{fig:tsne_chem}
    \vspace{-0.25in}
\end{figure}

\vspace{-0.1in}
Evidence from both unsupervised and supervised machine learning (ML) algorithms points to clear, distinctive structure in real-world networks from different domains. Dimensionality reduction using t-distributed stochastic neighbor embedding (t-SNE) shows clear clusters of graphs (see Figure~\ref{fig:tsne_chem}. Specifically, Facebook, Cheminformatics, Retweet, Brain, Social and Web/Technological graphs are able to be identified both visually and using k-means clustering.
\vspace{-0.1in}

\begin{figure}[t!]
    \vspace{-8mm}
    \centering
    \includegraphics[width=0.75\linewidth]{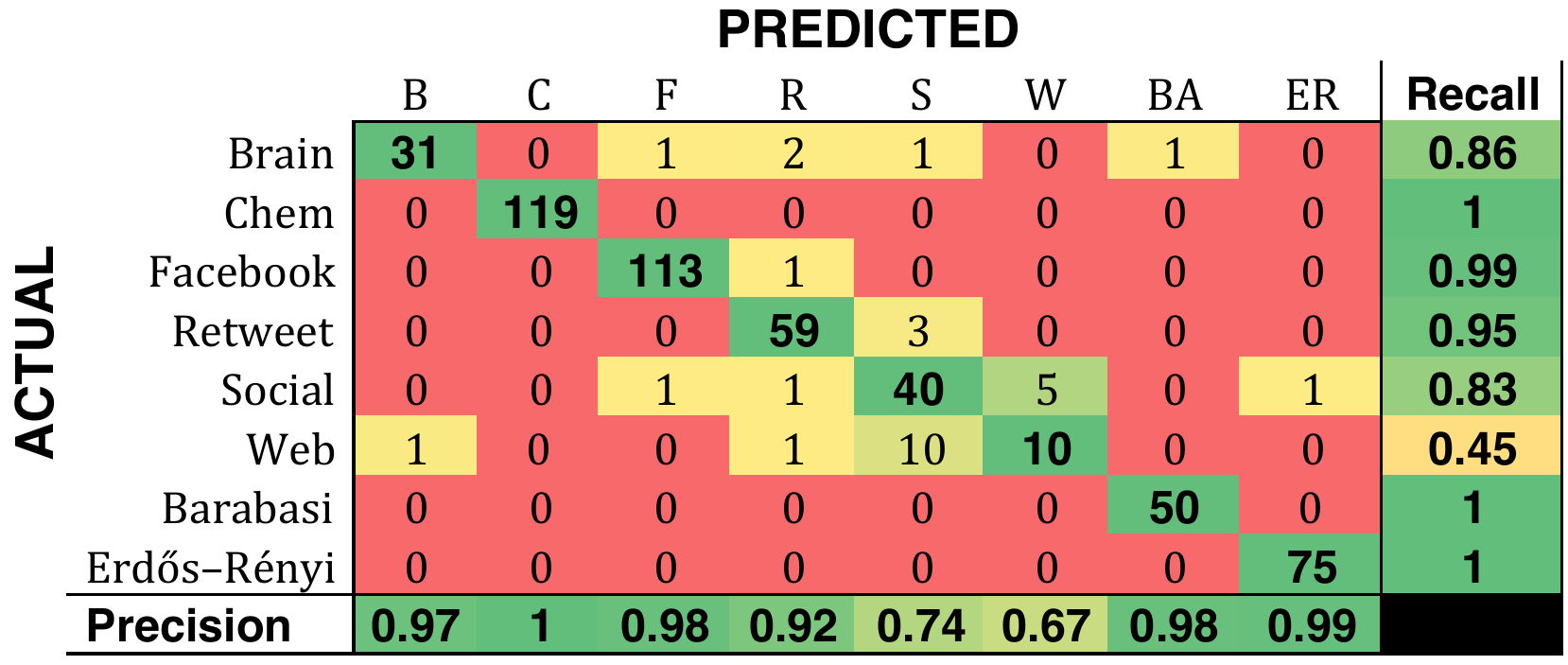}
    \vspace{-2mm}
    \caption{Contigency Matrix for Classification from a Random Forest Model}
    \label{fig:classification-results}
    \vspace{-0.15in}
\end{figure}

\normalsize
Similarly, standard classification algorithms are able to accurately classify graphs from each of those categories. 
A summary of the classification results are shown in Figure~\ref{fig:classification-results} and supports several important findings. 
First, we see that even though Erd\H{o}s-Reny\'i (ER) and Barabasi graphs (BA) are intended to model real-networks, they are distinct enough from their inspirations that only two other networks are classified as either BA or ER. This result questions the efficacy of testing algorithms/ideas intended for real networks on synthetic models. Second, it is apparent that graphs normally labeled ``Web'' are difficult to distinguish from social graphs. Deeper evaluation reveals that several of the web graphs represent pages within a specific social community, which could therefore influence the network's structure. Likewise, several social graphs are from very techno-centric realms and could be reasonably labeled as a web graph.

Additional tests show that two categories of graphs initially excluded due to low instances could be combined with existing categories with a minimal loss in accuracy. Results from k-means clustering indicate that Web and Technological graphs as well as Brain and Biological networks have similar properties. 
Finally, careful analysis of the mislabeled graphs in Figure~\ref{fig:classification-results} provides interesting network/category specific findings and suggestions. For example, 10 of the 36 brain networks are non-human, however all 5 graphs that are mislabeled are non-human. This is strong evidence that either the human networks are truly distinct from the non-humans, or the network discovery process is not sufficiently standardized for neuro-networks. Also interesting was that a visual inspection of the graph mislabeled as a retweet network shows surprising similarities. 
This suggests that using classification models provide valuable insight into alternative research techniques for crossing disciplines.

\vspace{-0.1in}
\section{Conclusions}
\vspace{-0.1in}
This work makes two important findings.
First, real-world networks from various domains have distinct structural properties that allow us to predict with high accuracy the category of an arbitrary network.
Second, classifying synthetic networks is trivial as our models can easily distinguish between synthetic graphs and the real-world networks they are supposed to model.

\vspace{-0.15in}
\bibliographystyle{splncs}
\bibliography{bibfile} %print bibliography

\end{document}